\documentclass[11pt]{article}
\usepackage{epsfig}
\usepackage{latexsym}
\usepackage{amsmath}
\usepackage{amssymb}

\newtheorem{theorem}{Theorem}{\bf}{\it}
{\bf}{\rm}
{\bf}{\it}
{\bf}{\it}
\newtheorem{lemma}[theorem]{Lemma}{\bf}{\it}
\renewcommand{\forall}{\mbox{for all}\,\,}

 \def\C{{\mathfrak C}} \def\mB{{\mathbb B}}
\def\D{{\cal D}} \def\G{{\bf G}} \def\H{{\cal H}}
 \def\mc{{\mathfrak
    c}}\def\md{{\mathfrak d}} \def\munt{\underline m}
\def\nunt{\underline n} \def\M{{\bf M}} 
\def\h{e_0^\perp} \def\fphi{\Phi} \def\W{{\cal W}}
\def\phi{\varphi}

\def\angle{\sphericalangle} \def\Ad{{\rm{Ad}}} \def\Aut{{\rm Aut}}
\def\ba{{\bf a}} \def\bF{{\bf F}} \def\complex{{\mathbb C}}
 \def\epsilon{\varepsilon}
\def\Halmos{\quad\hfill$\Box$}  
  
  \def\supp{{\rm supp}}
 
\def\test{{\mathfrak D}}\def\skippie{\smallskip\smallskip}

\def\integers{{\mathbb Z}} \def\complex{{\mathbb C}}
\def\reals{{\mathbb R}}

        \title{Spin, Statistics, and Reflections\\
          I. Rotation Invariance}
        \author{Bernd Kuckert\\
          II. Institut f\"ur Theoretische Physik\\ Luruper Chaussee
          149, 22761 Hamburg, Germany}

\begin{document}\maketitle
\begin{abstract}
  The universal covering of $SO(3)$ is modelled as a reflection group
  $\G_R$ in a representation independent fashion.  For relativistic
  quantum fields, the Unruh effect of vacuum states is known to imply
  an intrinsic form of reflection symmetry, which is referred to as
  {\it modular P$_1$CT-symmetry} \cite{BW75,BW76,GL95}. This symmetry
  is used to construct a representation of $\G_R$ by pairs of modular
  P$_1$CT-operators. The representation thus obtained satisfies
  Pauli's spin-statistics relation.
\end{abstract}

\section{Introduction}

A vacuum state of a quantum field theory usually exhibits the Unruh
effect, i.e., a uniformly accelerated observer experiences it as a
thermal state whose temperature is proportional to his acceleration
\cite{Unr76}.  This has been shown by Bisognano and Wichmann
\cite{BW75,BW76} for finite-component quantum fields (in the Wightman
setting). For general quantum fields, it has recently been derived
from the mere condition that each vacuum state exhibits passivity to
each inertial or uniformly accelerated observer \cite{Kuc02}, i.e.,
that in the observer's rest frame, no engine can extract energy from
the state by cyclic processes.\footnote{Two related uniqueness results
  can be found in Refs. \ref{Kuc97} and \ref{Kuc01}.}

By the theorem of Bisognano and Wichmann mentioned above, all familiar
quantum fields also exhibit an intrinsic form of
PCT-symmetry.\footnote{cf.  also Refs. \ref{GL95}, \ref{Kuc97}, and
  \ref{Kuc01}.} Namely, one can assign to each Rindler wedge $\W$,
i.e., the set $\W_1:=\{x_1\geq|x_0|\}$ or its image under some
Poincar\'e transformation, an antiunitary involution $J_\W$. This
assignment is an intrinsic construction using the vacuum vector and
the field operators only. It is also basic to the so-called modular
theory due to Tomita and Takesaki, where an operator like $J_\W$ is
called a {\it modular conjugation}.  $J_\W$ then implements a {\it
  P$_1$CT-symmetry}, i.e., a linear reflection in charge and at the
edge of $\W$. This property is called {\it modular P$_1$CT-symmetry}.
Note as an aside that this symmetry is a typical property of
$1+2$-dimensional quantum fields as well, whereas these fields do not
exhibit PCT-symmetry as a whole \cite{Man69}.

Modular P$_1$CT-symmetry is a consequence of the Unruh effect
\cite{GL95}, but the converse implication does not hold: There are
examples of P$_1$CT-symmetric quantum fields that do not exhibit the
Unruh property \cite{BDFS}.

Guido and Longo have derived Pauli's spin-statistics relation from the
Unruh effect for general quantum fields in $1+3$ dimensions
\cite{GL95}.\footnote{cf. also Refs. \ref{FM93}, \ref{GL96}, and
  \ref{GLRV}.} Independently from this, the present author derived the
spin-statistics relation making use of modular P$_1$CT-symmetry only
\cite{Kuc95}.  

This symmetry was assumed for the field's observables only, but since
use of a theorem due to Doplicher and Roberts \cite{DR90} was made
later on, the result of Ref. \ref{Kuc95} is confined to the
massive-particle excitations of the vacuum.

In Ref. \ref{GL95} the Unruh effect was assumed for the whole field on
the one hand. On the other hand, no use of the Doplicher-Roberts
theorem was made, so a much larger class of fields and states was
included; even fields that are covariant with respect to {\it more
  than one} representation of the universal covering group
of $L_+^\uparrow$, among which there may be both
representations satisfying and violating Pauli's relation
\cite{Str67}. What one did obtain was a unique representation
satisfying the Unruh effect. This representation exhibits Pauli's
spin-statistics connection. All spin-statistics theorems obtained
before did not admit this extent of generality.

This paper is the first of two that generalize the result of Ref.
\ref{Kuc95} in this spirit as well. Assuming P$_1$CT-symmetry with
respect to all Rindler wedges whose edges are two-dimensional planes
in a given tim-zero plane, a covariant unitary representation $\tilde
W$ of the rotation group's universal covering is constructed. This
representation satisfies Pauli's spin-statistics relation. The
argument does not make use of the Doplicher-Roberts theorem and
applies to general relativistic quantum fields.

Like its predecessor in Ref. \ref{Kuc95}, the argument is crucially
based on the fact that each rotation in $\reals^3$ can be implemented
by combining two reflections at planes.  This is, as such, well known
for both $SO(3)$ and $L_+^\uparrow$. A corresponding result for the
universal coverings of these groups is, however, less elementary to
obtain.

In Section \ref{SU2}, a model $\G_R\cong SU(2)$ of the universal
covering group $\widetilde{SO(3)}$ of $SO(3)$ will be constructed from
nothing except pairs of ``reflections along normal vectors'', i.e.,
from the family $(j_a)_{a\in S^2}$, where $j_a$ is the reflection at
the plane $a^\perp$. This representation-independent construction is
set up according to the needs of the spin-statistics theorem to be
proved later on. A model $\G_L\cong SL(2,\complex)$ of
$\widetilde{L_+^\uparrow}$ will be constructed in a forthcoming paper.
It is to be expected that the universal coverings of other Lie groups
could be constructed the same way.

Recently it has been shown by Buchholz, Dreyer, Florig, and Summers
that this structure has a representation theoretic consequence:
unitary representations of $L_+^\uparrow$ can be constructed from a
system of reflections satisfying a minimum of covariance conditions,
as they are satisfied by the modular conjugations of a quantum field
with modular P$_1$CT-symmetry \cite{BDFS,Flo99,BS}. This raises the
question how to generalize these results to $\G_R$ and $\G_L$, the
goal being a considerable generalization of the spin-statistics
analysis in Ref.  \ref{Kuc95}.

In Section \ref{spin and statistics}, it is shown that this can,
indeed, be accomplished for $\G_R$; the group $\G_L$ will be treated
in the forthcoming paper. If a quantum field exhibits modular
P$_1$CT-symmetry, then it is elementary to build a distinguished
representation $\tilde W$ of $\G_R$ from the modular conjugations that
implement P$_1$CT-symmetry.  This representation can, eventually,
easily be shown to conform with Pauli's spin-statistics principle.

It is well known that not all $\G_R$-covariant quantum fields exhibit
the spin-statistics relation, and it should be remarked that even for
Lorentz covariant fields there are counterexamples \cite{Str67}.  This
means that some condition specifying the representation or field under
consideration is needed for whatever spin-statistics theorem. In the
early spin-statistics theorems, this condition was that the number of
internal degrees of freedom is finite, in this paper the condition is
that the representation is constructed from modular P$_1$CT-operators.
At the moment, such sufficient conditions are all one has in the
relativistic setting; only in the setting of nonrelativistic quantum
mechanics, a both sufficient and necessary condition has been
established \cite{Kuc04,KM04}.

\section{$\widetilde{SO(3)}$ as a reflection group}
\label{SU2}

There are many ways to model the universal covering group of the
rotation group $SO(3)=:R$. Among topologists, ``the'' universal
covering group is the group $\widetilde{SO(3)}$ of homotopy classes of
curves starting at some base point, physicists are more familiar with
$SU(2)$, but these are, of course, not the only examples of simply
connected covering groups. As a new model, a group $\G_R$ will be
constructed in this section from pairs of ``reflections along normal
vectors'', i.e., from the family $(j_a)_{a\in S^2}$, where $j_a$ is
the reflection at the plane $a^\perp$.

\bigskip Let ${\M}_{R}$ be the pair groupoid of $S^2$, i.e., the set
$S^2\times S^2$ endowed with the concatenation
$(a,b)\circ(b,c):=(a,c)$.  Then the map $\rho:\,\M_R\to R$ defined by
$\rho(a,b):=j_aj_b$ is well known to be surjective. Namely,
$\rho(a,a)=1$ for all $a\in S^2$. For $\sigma\neq 1$, choose $\tau\in
R$ such that $\tau^2=\sigma$; if $a\in S^2$ is perpendicular to the
axis of $\sigma$, then $\rho(\tau a,a)=\sigma$.

Call $(a,b)$ and $(c,d)$ equivalent if $\rho(a,b)=\rho(c,d)$ and if
there exists a $\sigma\in R$ commuting with $\rho(a,b)$ and satisfying
$(a,b)=(\sigma^2c,\sigma^2d)$.\,\footnote{The square superscripts are
  missing in earlier versions, including the published one. Without
  them, the assumption $\sigma\rho(a,b)\sigma^{-1}=\rho(a,b)$ would be
  redundant, and $(a,b)\sim(a,-b)$ for $a\perp b$ (let $\sigma$ be the
  rotation by the angle $\pi$ around the axis $a\reals$).} Let
${\G}_{R}$ be the quotient space ${\M}_{R}/\!\!\sim$ associated with
this equivalence relation, and let $\pi:\,{\M}_{R}\to {\G}_{R}$ denote
the corresponding canonical projection. Define $\tilde\rho:\,\G_R\to
R$ by $\tilde\rho(\pi(\munt )):=\rho(\munt )$ for all $\munt\in\M_R$.
Then the diagram
\begin{equation}\label{rhos}
\begin{array}{ccc}
{\M}_{R}&\stackrel{\pi}{\longrightarrow}&{\G}_{R} \\ \\
\rho\downarrow& \swarrow\tilde\rho&\\ \\
R& &
\end{array}
\end{equation}
commutes by construction. All maps in this diagram are continuous:
$\pi$ is continuous by definition, and continuity of $\rho$ is
elementary to show. The proof for $\tilde\rho$ is elementary as well:
given any open set $M\subset R$, the pre-image $\tilde\rho^{-1}(M)$ is
open if and only if $\pi^{-1}(\tilde\rho^{-1}(M))$ is open. This set
coincides with $\rho^{-1}(M)$, which is open by continuity of $\rho$.

Defining $\pm1:=\pi(a,\pm a)$ for arbitrary $a\in S^2$, and
$-\pi(a,b):=\pi(a,-b)$ for $(a,b)\in\M_R$, one verifies that
$\tilde\rho^{-1}(\sigma)$ consists of two equivalence classes for
each $\sigma\in R$.

\begin{lemma}\label{lem:rho tilde}
(i) $\G_R$ is a Hausdorff space.

(ii) $\tilde\rho$ is a two-sheeted covering map.
\end{lemma}
Before proving this lemma, we introduce some notation.

\smallskip\smallskip {\it Notation.} Denote the set $R\backslash\{1\}$
by $\dot R$. For each $\sigma\in\dot R$, let $A(\sigma)$ be the
rotation axis of $\sigma$.  If $\ba\in A(\sigma)$ is one of the two
unit vectors in $A(\sigma)$, then there is a unique
$\alpha\in(0,2\pi)$ such that $\sigma$ is the right-handed rotation
around $\ba$ by the angle $\alpha$. The vector $\ba$ and the angle
$\alpha$ determine $\sigma$, and occasionally we use the notation
$[\ba,\alpha]$ for $\sigma$. Note that
$[\ba,\alpha]=[-\ba,2\pi-\alpha]$.

Denote the set $\rho^{-1}(\dot R)$ by $\dot\M_R$. To each
$(a,b)\in\dot\M_R$, assign the axial unit vector
$\ba(a,b):=\frac{a\times b}{|a\times b|}$, and denote by
$\angle(a,b)\in(0,\pi)$ the angle between $a$ and $b$.  Note that
$\rho(\underline m)=[\ba(\underline m),2\angle(\underline m)]$ for all
$\munt\in\dot\M_R$.

Denote the set $\tilde\rho^{-1}(\dot R)$ by $\dot\G_R$. Since
$\munt\sim\nunt$ implies $\ba(\underline m)=\ba(\underline n)$ and
$\angle(\munt)=\angle(\nunt)$ one can define
$\tilde\ba(\pi(\munt)):=\ba(\munt)$ and
$\tilde\angle(\pi(\munt)):=\angle(\munt)$.  Note that
$\tilde\rho(g)=[\tilde\ba(g),2\tilde\angle(g)]$ for all
$g\in\dot\G_R$.

\skippie {\it Proof of Lemma \ref{lem:rho tilde}.(i).}
Define $\dot\mB_\pi:=\{x\in\reals^3:\,|x|\in(0,\pi)\}$, and assign to
each $x\in\dot\mB_\pi$ the rotation
$\tau(x):=\left[x/|x|,\,|x|\right]$.  Choose any $x\in\dot\mB_\pi$ and
an $a\in S^2\cap x^\perp$, and put
$\xi_a(x):=\pi(\tau(x)a,a)\in\dot\G_R$. One then obtains
$\xi_a(x)=\xi_b(x)$ for all $b\in S^2\cap x^\perp$, so a map
$\xi:\,\dot\mB_\pi\to\dot\G_R$ is well defined by $\xi(x):=\xi_a(x)$,
where $a\in S^2\cap x^\perp$ is arbitrary.

$\xi$ is inverse to the map $\eta:\,\dot\G_R\to\dot\mB_\pi$ defined by
$\eta(g):=-\tilde\angle(g)\tilde\ba(g)$. Namely, since $b\perp a\times b$
for all $(a,b)\in\dot\M_R$, one has 
\begin{align*}
  \xi(\eta(\pi(a,b)))&=\xi\left(\angle(a,b)\cdot\ba(a,b)\right)
  =\pi\left(\,\tau\left(-\angle(a,b)\ba(a,b)\right)\,b,b\right)\\
&=\pi\left(\left[-\ba(a,b),\angle(a,b)\right]b,b\right)
=\pi(a,b).
\end{align*}
So $\eta$ is continuous, surjective, and has a continuous inverse, so
$\eta$ is a homeomorphism, and $\dot\G_R$ is a Hausdorff space.

It remains to construct disjoint neighborhoods of two distinct points
$g, h\in\G_R$ for the case that $g=\pm1$ and $h\in\G_R$ is arbitrary.

If $g=1$, then $\tilde\angle(h)\neq0$, so there exist disjoint open
neighborhoods $X$ and $Y$ of $0$ and $\tilde\angle(h)$ in the
topological space $[0,\pi]$, respectively. Since the map
$\tilde\angle$ is continuous, the sets $U:=\tilde\angle^{-1}(X)$ and
$V:=\tilde\angle^{-1}(Y)$ are disjoint neighborhoods of $1$ and $h$.
If $g=-1$, there exist disjoint neighborhoods $U'$ and $V'$ of $-g$
and $-h$, so $-U'$ and $-V'$ are disjoint neighborhoods of $g$ and
$h$, respectively.

{\it Proof of (ii).}
Define $\hat\rho:\,\dot\mB_\pi\to\dot R$ by
$\hat\rho(x):=[x/|x|,2|x|]$. Then the diagram 
\begin{equation}
\begin{array}{ccc}
 & &\dot{\G}_{R} \\ \\
 & \tilde\rho|_{\dot\G_R}\swarrow\hspace{0.5cm}&\downarrow\eta\\ \\
\dot R&\stackrel{\hat\rho}{\longleftarrow}&\dot\mB_\pi
\end{array}
\end{equation}
commutes. $\hat\rho$ is a two-sheeted covering map, and $\eta$
is a homeomorphism, so $\tilde\rho|_{\dot\G_R}=\hat\rho\circ\eta$ is a
two-sheeted covering map.

In order to prove that $\tilde\rho$ as a whole is a covering map,
it remains to be shown that $\tilde\rho$ is open not only on
$\dot\G_R$, but also in $\pm1$. Since $\G_R$ is Hausdorff, since
$\dot\G_R$ is a two-sheeted covering space of $\dot R$, and since
$\tilde\rho^{-1}(1)=\{\pm1\}$ contains, like all other fibers of
$\tilde\rho$, precisely two elements, it then follows that
$\tilde\rho$ has continuous local inverses everywhere.

So let $(\sigma_n)_n$ be any sequence in $\dot R$
converging to $1$, then some sequence $(g_n)_n$ in $\dot\G_R$ needs to
be found with $\tilde\rho(g_n)=\sigma_n$ for all $n$ and $g_n\to1$;
note that $(-g_n)_n$ then satisfies $\tilde\rho(-g_n)=\sigma_n$ as
well and converges to $-1$.
  
For each $g\in\dot\G_R$, one has $\tilde\angle(g)\leq\pi/2$ or
$\tilde\angle(-g)\leq\pi/2$. It follows that for each $n$ some
$g_n\in\tilde\rho^{-1}(\sigma_n)$ can be chosen such that
$\tilde\angle(g_n)\leq\pi/2$. Since $[0,\pi/2]$ is compact, the
sequence $(\tilde\angle(g_n))_n$ has at least one accumulation point,
and since $\sigma_n$ tends to $1$, the only possible accumulation
point in the interval $[0,\pi/2]$ is zero. It follows that
$\tilde\angle(g_n)$ tends to zero and, hence, that $g_n$ tends to $1$,
proving that $\tilde\rho$ is open. \Halmos

\bigskip The reason why this proof is nontrivial is that $\rho$ and
$\pi$ are not open. If this were the case, $\G_R$ would directly
inherit the Hausdorff property from $\M_R$, and the proof that
$\tilde\rho$ is a covering map would be elementary. But neither $\rho$
nor $\pi$ is open.

In order to see this, let $(\sigma_n)_n$ be any sequence of rotations
around some fixed $a\in S^2$, and suppose this sequence to converge to
$1$. If $\rho$ were open, one would have to find, for each $\underline
m\in\pi^{-1}(1)$ a sequence $(\underline m_n)_n$ converging to
$\underline m$ and satisfying $\rho(\underline m_n)=\sigma_n$ for all
$n$. Now choose $\underline m=(a,a)$. Since $a\in A(\sigma_n)$ for all
$n$, one knows for all $(b_n,c_n)\in\rho^{-1}(\sigma_n)$ that both
$b_n$ and $c_n$ are perpendicular to $a$. As a consequence, no
sequence $(\underline m_n)_n$ with $\rho(\underline m_n)=\sigma_n$ for
all $n$ can coverge to $\underline m=(a,a)$.

$\pi$ cannot be open either, since this would, by diagram \ref{rhos}
and the preceding Lemma, imply that $\rho$ is open. Only the
restrictions of $\rho$ and $\pi$ to $\rho^{-1}(\dot R)$ are open.

\begin{theorem}\label{theorem R}

(i) ${\G}_{R}$ is simply connected.

(ii) There is a unique group product 
  $\odot$ on ${\G}_{R}$ such that the diagram
\begin{equation}\label{diagram G_R}
\begin{array}{ccc} {\M}_{R}\times {\M}_{R}
  &\stackrel{\circ}{\longrightarrow}&{\M}_{R}\\ \\
  \downarrow\pi\times\pi& &\downarrow\pi\\ \\
  {\G}_{R}\times {\G}_{R}&\stackrel{\odot}{\longrightarrow}&{\G}_{R}\\ \\
  \downarrow\tilde\rho\times\tilde\rho& &\downarrow\tilde\rho\\ \\
  R\times R&\stackrel{\cdot}{\longrightarrow}&R
\end{array}
\end{equation}
commutes.
\end{theorem}

{\it Proof of (i).} $\G_R=\pi(\M_R)$ is pathwise connected because
$\M_R=S^2\times S^2$ and because $\pi$ is continuous. Together with 
Lemma \ref{lem:rho tilde}, this implies the statement, since the
fundamental group of $R$ is ${\mathbb Z}_2$.

\skippie
{\it Proof of (ii).} The outer arrows of the diagram commute, so it
suffices to prove the existence and uniqueness of a group product
conforming with the lower part. But it is well known that each simply
connected covering space $\tilde G$ of an arbitrary topological group
$G$ can be endowed with a unique group product $\odot$ such that $G$
is a covering group.\footnote{See, e.g., Props. 5 and 6
  in Sect. I.VIII. in Ref. \ref{Chev}.}\Halmos

\section{Spin \& Statistics}\label{spin and statistics}

The preceding section has provided the basis of a general
spin-statistics theorem, which is the subject of this section. From an
intrinsic form of symmetry under a charge conjugation combined with a
time inversion and the reflection in {\it one} spatial direction,
which is referred to as {\it modular P$_1$CT-symmetry}, a strongly
continuous unitary representation $\tilde W$ of $\G_R$ will be
constructed using the above and related reasoning. It is, then,
elementary to show that $\tilde W$ exhibits Pauli's spin-statistics
relation.

\smallskip\smallskip 
In order to make the notion of rotation meaningful, fix a
distinguished time direction by choosing a future-directed timelike
unit vector $e_0$.  The $2$-sphere of unit vectors in the time-zero
plane $\h$ will be called $S^2$.

Let $F$ be an arbitrary quantum field on $\reals^{1+3}$ in a Hilbert
space $\H$. The following standard properties of relativistic quantum
fields will be used here.

\begin{itemize}
\item[(A)] {\it Algebra of field operators.} Let $\C$ be a linear
  space of arbitrary dimension,\footnote{$\C$ is the
    ``component space'', and its dimension equals the number of
    components, which may be infinite in what follows.} and denote by
  $\test$ the space $C_0^\infty(\reals^{1+3})$ of test functions on
  $\reals^{1+3}$. The field $F$ is a linear function that assigns to
  each $\fphi\in\C\otimes\test$ a linear operator $F(\fphi)$ in a
  separable Hilbert space $\H$. 
\begin{itemize}
\item[(A.1)] $F$ is free from redundancies in $\C$, i.e., if
  $\mc,\md\in\C$ and if $F(\mc\otimes\phi))=F(\md\otimes\phi)$ for
  all $\phi\in\test$, then $\mc=\md$.
\item[(A.2)]  Each field operator $F(\fphi)$ and its
  adjoint $F(\fphi)^\dagger$ are densely defined. There exists a dense
  subspace $\D$ of $\H$ contained in the domains of $F(\fphi)$ and
  $F(\fphi)^\dagger$ and satisfying $F(\fphi)\D\subset\D$ and
  $F(\fphi)^\dagger\D\subset\D$ for all $\fphi\in\C\otimes\test$.
\end{itemize}  
Denote by ${\bf F}$ the algebra generated by all $F(\fphi)|_\D$ and
  all $F(\fphi)^\dagger|_\D$.  Defining an involution $*$ on ${\bf F}$
  by $A^*:=A^\dagger|_\D$, the algebra ${\bf F}$ is endowed with the
  structure of a $*$-algebra.

For each $a\in S^2$, denote by
$\W_{a}:=\{x\in\reals^{1+3}:\,xa>|xe_0|\}$ the {\it Rindler wedge}
associated with $a$,\footnote{An observer who is uniformly accelerated
in the direction $a$ can interact with precisely the events in
$\W_{a}$.}  and let $\bF({a})$ be the algebra generated by all
$F(\mc\otimes\varphi)|_\D$ and all $F(\mc\otimes\varphi)^\dagger|_\D$ with
$\supp(\varphi)\subset\W_a$. The algebra $\bF(a)$ inherits the
structure of a $*$-algebra from $\bF$ by restriction of $*$. 
\begin{itemize}
\item[(A.3)] $\bF(a)$ is
nonabelian for each $a$, and $a\neq b$ implies $\bF(a)\neq\bF(b)$.
\end{itemize}

\item[(B)] {\it Cyclic vacuum vector.} There exists a vector
  $\Omega\in\H$ that is 
  cyclic with respect to each $\bF(a)$.

\item[(C)] {\it Normal commutation relations.} There exists a unitary
   and self-adjoint operator ${k}$ on $\H$ with $k\Omega=\Omega$ and
   with $k\bF({a})k=\bF({a})$ for all $a\in S^2$. Define
   $F_\pm:=\frac{1}{2}(F\pm{k} F{k})$.  If $\mc$ and $\md$ are
   arbitrary elements of $\C$ and if
   $\varphi,\psi\in\test$ have spacelike separated
   supports, then
\begin{align*}
  F_+(\mc\otimes\varphi)F_+(\md\otimes\psi)&=F_+(\md\otimes\psi)F_+(\mc\otimes\varphi),\\
  F_+(\mc\otimes\phi)F_-(\md\otimes\psi)
  &=F_-(\md\otimes\psi)F_+(\mc\otimes\varphi),\quad{\rm and}\\
  F_-(\mc\otimes\varphi)F_-(\md\otimes\psi)&=-F_-(\md\otimes\psi)F_-(\mc\otimes\varphi).
\end{align*}
\end{itemize}
The involution ${k}$ is the {\it statistics operator}, and $F_\pm$ are
the bosonic and fermionic components of $F$, respectively. Defining
$\kappa:=(1+ik)/(1+i)$ and $F^t(\md\otimes\psi):=\kappa
F(\md\otimes\psi)\kappa^\dagger$, the normal commutation relations
read
$$[F(\mc\otimes{\varphi}),F^t(\md\otimes\psi)]=0.$$
This property is referred
to as {\it twisted locality}. Denote
$\bF(a)^t:=\kappa\bF(a)\kappa^\dagger$.

\smallskip\smallskip These properties imply that $\Omega$ is {\it
  separating} with respect to each algebra $\bF({a})$, i.e., for each
$A\in\bF(a)$, the condition $A\Omega=0$ implies $A=0$.\footnote{If
  $A\Omega=0$ and $B,C\in\bF(-a)^t$, then $0=\langle
  BC\Omega,A\Omega\rangle= \langle C\Omega,A B^*\Omega\rangle$, so
  $A=0$ by cyclicity of $\Omega$.}

As a consequence, an antilinear operator
$R_{a}:\,\bF({a})\Omega\to\bF({a})\Omega$ is defined by
$R_{a}A\Omega:=A^*\Omega$. This operator is closable.\footnote{By
  twisted locality, the operator $\kappa R_{-a}\kappa^\dagger$ is
  formally adjoint to $R_a$. Namely, if
  $A\in\kappa\bF(-a)\kappa^\dagger$ and $B\in\bF(a)$, then $\langle
  A\Omega,R_aB\Omega\rangle=\langle A\Omega,B^*\Omega\rangle=\langle
  B\Omega,A^*\Omega\rangle =\langle B\Omega,\kappa
  R_{-a}\kappa^\dagger A\Omega\rangle$. Since $\kappa
  R_{-a}\kappa^\dagger$ is densely defined, it follows that $R_a$ is
  closable.} Its closed extension $S_{a}$ has a unique polar
decomposition $S_{a}=J_{a}\Delta_{a}^{1/2}$ into an antiunitary
operator $J_{a}$, which is called the {\it modular conjugation}, and a
positive operator $\Delta_{a}^{1/2}$, which is called the {\it modular
  operator}. $J_a$ is an involution.\footnote{$R_a^2=1$ implies
  $S_a^2=1$, so $J_a\Delta_a^{1/2}=S_a=S_a^{-1}=\Delta_a^{-1/2}J_a^*$,
  i.e., $J_a^2\Delta_a^{1/2}=J_a\Delta_a^{-1/2}J_a^*$. Since
  $J_a\Delta_a^{-1/2}J_a^*$ is positive, one obtains $J_a^2=1$ and
  $J_a\Delta^{-1/2}J_a=\Delta^{1/2}$ from the uniqueness of the polar
  decomposition \cite{BR}.}  $S_{a}$, $J_{a}$, and $\Delta^{1/2}_{a}$
are the objects of the so-called {\it modular theory} developed by
Tomita and Takesaki.\footnote{The original work \cite{Tak70} directly
  applies to von-Neumann algebras, which are normed. But also for the
  present setting this structure has been applied earlier, e.g., in
  the classical papers of Bisognano and Wichmann \cite{BW75,BW76}.
  See, also, Ref. \ref{Inoue} for a monograph on the Tomita-Takesaki
  theory of unbounded-operator algebras.}

\smallskip\smallskip
For each $a\in S^2$, let $j_a$ be
  the orthogonal reflection at the plane $a^\perp\cap
  e_0^\perp$,\footnote{i.e., the linear reflection with $j_aa=-a,\quad
  j_ae_0=-e_0$, and $j_ax=x$ for all $x\in a^\perp\cap e_0^\perp$.}
  and for each $\phi\in\test$, define the test function
  $j_a\phi\in\test$ by $j_a\phi(x):=\phi(j_ax)$.

\begin{itemize}
\item[(D)]{\it Modular P$_1$CT-symmetry.} For each $a\in S^2$, there
  exists an antilinear involution $C_a$ in $\C$ such that for all
  $\mc\in\C$ and $\varphi\in\test$, one has
  $$J_{a}F(\mc\otimes\varphi)J_{a}=F^t(C_a\mc\otimes
  \overline{j_a\varphi}).$$
  The map $a\mapsto J_a$ is strongly
  continuous.\footnote{If one assumes covariance with respect to some
    strongly continuous representation of $G_R$ (which may also
    violate the spin-statistics connection), this is straightforward
    to derive; cf.  Lemma \ref{lem:com rel}. But covariance, as such,
    is not needed.}
\end{itemize}
It will now be shown that pairs of modular P$_1$CT-reflections give rise to
a strongly continuous representation of $\G_R$ which exhibits Pauli's
spin-statistics connection.
\begin{lemma}\label{lem:com rel}
  Let $K$ be a unitary or antiunitary operator in $\H$
  such that $K\D= \D$ and $K\Omega=\Omega$, and suppose there are
  $a,b\in S^2$ such that $K\bF({a})K^\dagger=\bF({b})$. Then
  $KJ_{a}K^\dagger=J_{b}$, and $K\Delta_a K^\dagger=\Delta_b$.
\end{lemma}
{\it Proof.} If $A\in\bF(b)$, then $KS_{a}K^\dagger
A\Omega=KS_{a}\underbrace{K^\dagger AK}_{\in\bF({a})}\Omega
=A^*\Omega=S_{b}A\Omega$. The statement now follows by the uniqueness
of the polar decomposition.\Halmos

\bigskip
In particular, this lemma implies
\begin{equation}\label{kJk}
kJ_ak=J_a,\quad{\rm whence}\quad J_a\kappa=\kappa^\dagger J_a
\end{equation}
by definition of $k$. Using twisted locality, the lemma also implies
\begin{equation}\label{minus a}
\kappa J_{a}\kappa^\dagger\,=\kappa^\dagger J_a\kappa=J_{-a}
\end{equation}
which, in turn, implies
\begin{equation}\label{reflections}
J_{a}J_{b}J_{a}=J_{-j_{a}{b}}=J_{j_aj_bb}=J_{\rho(a,b)b}
\end{equation}
by modular P$_1$CT-symmetry.

Define a map $W$ from $\M_R$ into the unitary group of $\H$
by $W(a,b):=J_aJ_b$.
\begin{lemma}\label{W tilde}

(i) $\munt \sim \nunt $ implies $W(\munt )=W(\nunt )$.

(ii) $W(\munt)=W(\nunt)$ implies $\rho(\munt)=\rho(\nunt)$.

\end{lemma}

{\it Proof of (i).} The proof of Lemma 2.4 in Ref. \ref{BS} can be
taken without any relevant changes. Despite the fact that the
Buchholz-Summers paper is confined to bosonic fields, which, in
particular, implies $J_a=J_{-a}$, it is straightforward to translate
their proof to the present setting. This will not be spelled out here.
The proof makes use of the continuous dependence of $J_a$ from $a$ assumed
in Assumption (D).

\smallskip
{\it Proof of (ii).} $\rho(\munt)\neq\rho(\nunt)$ would imply that
there is some $b\in S^2$ such that $\rho(\munt)b\neq\rho(\nunt)b$,
so $\bF(\rho(\munt)b)\neq\bF(\rho(\nunt)b)$ by Assumption (A), whence
$W(\munt)\bF(b)W(\munt)^*\neq W(\nunt)\bF(b)W(\nunt)^*$ by Assumption (D),
i.e., $W(\munt)\neq W(\nunt)$.\Halmos

\bigskip By this lemma, a map $\tilde W:\,{\G}_R\to
W({\M}_{R})$ is defined by $\tilde W(\pi(\munt)):=W(\munt)$, and another
map $\rho_W:\,W(\M_R)\to R$ is defined by
$\rho_W(W(\munt))=\rho(\munt)$.  The diagrams
\begin{equation}\label{diagram W}
(A)\\\quad\begin{array}{ccc}
{\M}_{R}&\stackrel{\pi}\longrightarrow&{\G}_R\\ \\
\rho\downarrow&\stackrel{W}{\searrow}&\downarrow\tilde W\\ \\
{R}&\stackrel{\rho_W}{\longleftarrow}&W({\M}_{R})
\end{array}
\quad{\rm and}\quad(B)\\\quad
\begin{array}{ccc}
{\M}_{R}&\stackrel{\pi}\longrightarrow&{\G}_R\\ \\
\rho\downarrow&\stackrel{\tilde\rho}{\swarrow}&\downarrow\tilde W\\ \\
{R}&\stackrel{\rho_W}{\longleftarrow}&W({\M}_{R})
\end{array}
\end{equation}
commute. 

\begin{theorem}

(i) There is a unique group product 
  $\odot_W$ on ${W({\M}_{R})}$ with the property that the diagram
\begin{equation}\label{diagram W_R}
\begin{array}{ccc} {\M}_{R}\times {\M}_{R}
  &\stackrel{\circ}{\longrightarrow}&{\M}_{R}\\ \\
  \downarrow \pi\times\pi & &\downarrow\pi\\ \\
  \G_R\times\G_R&\stackrel{\odot}{\longrightarrow}&\G_R\\ \\
  \downarrow\tilde W\times\tilde W& &\downarrow\tilde W\\ \\
  W({\M}_R)\times W({\M}_R)&\stackrel{\odot_W}{\longrightarrow}&W({\M}_R)\\ \\
  \downarrow\rho_W\times\rho_W& &\downarrow\rho_W\\ \\
  {R}\times {R}&\stackrel{\cdot}{\longrightarrow}&{R}
\end{array}
\end{equation}
commutes, i.e., $\tilde W$ is a homomorphism.

(ii) $\odot_W$ is the operator product in the algebra ${\cal B}(\H)$ of
bounded operators on $\H$, i.e., $\tilde W$ is a representation.

(iii) There is a representation $\tilde{D}$ of $G_R$ in $\C$ such that
\begin{equation}\label{modular covariance}
\tilde W(g)F(\mc\otimes\phi)\tilde W(g)^*=F(\tilde{D}(g)\mc\otimes\tilde\rho(g)\phi)\quad 
\forall g,\mc,\phi,
\end{equation}
where $\tilde\rho(g)\phi:=\phi(\tilde\rho(g)^{-1}\,\cdot)$.

\end{theorem}

\bigskip {\it Proof of (i).} The diagram already commutes if the
arrow representing $\odot_W$ is omitted.

For each $g\in\G_R$ and each $(a,b)\in\pi^{-1}(g)$ one has
$$\tilde W(\pm1)\tilde W(\pi(a,b))=W((\pm a,a)\circ(a,b))=W(\pm
a,b)=\tilde W(\pm(\pi(a,b))),$$
so $\tilde W(\pm1)\cong\integers_2$ 
if and only if $\tilde W(\pm g)\cong\integers_2$.

If $\tilde W(\pm 1)\cong\integers_2$, then $\tilde W$ is a bijection, so 
$\odot_W$ is defined by 
$$U\odot_WV:=\tilde W(\tilde W^{-1}(U)\odot\tilde W^{-1}(V)).$$

If $\tilde W(\pm1)\cong\{1\}$, then $\rho_W$ is a bijection, so $\odot_W$
is defined by 
$$U\odot_WV:=\rho_W^{-1}(\rho_W(U)\cdot\rho_W(V)).$$

{\it Proof of (ii).} The statement is nontrivial only on $\dot\G_R$.
Given $g,h\in\dot\G_R$, the planes $\tilde\ba(g)^\perp$ and
$\tilde\ba(h)^\perp$ intersect in an at least one-dimensional
subspace, so one can choose $(a,b)\in\pi^{-1}(g)$ and
$(c,d)\in\pi^{-1}(h)$ such that $b=c$ is in this intersection. Then
\begin{align*}
  \tilde W(\pi(a,b)\odot\pi(c,d))&=\tilde W(\pi((a,b)\circ(b,d)))\\
  &=\tilde W(\pi(a,d))=W(a,d)\\
  &=J_{a}J_{d}=J_{a}J_bJ_bJ_d\\
  &=W(a,b)W(b,d)=\tilde W(\pi(a,b))\tilde W(\pi(b,d))\\
  &=\tilde W(\pi(a,b))\tilde W(\pi(c,d)).
\end{align*}

{\it Proof of (iii).} Define a map $D$ from $\M_R$ into the
automorphism group $\Aut(\C)$ of $\C$ by $D(a,b):=C_aC_b$. If
$(a,b)\sim(c,d)$, then modular P$_1$CT-symmetry implies
\begin{align*}
F(C_aC_b\mc\otimes j_aj_b\phi)&=W(a,b)F(\mc\otimes\phi)W(a,b)^*\\
&=W(c,d)F(\mc\otimes\phi)W(c,d)^*\\&
=F(C_cC_d\mc\otimes j_cj_d\phi)\\&
=F(C_cC_d\mc\otimes j_aj_b\phi)
\end{align*}
for all $\mc$ and all $\phi$. Using Assumption (A.1), one obtains
$C_aC_b\mc=C_cC_d\mc$ for all $\mc$, so $D(a,b)=D(c,d)$, and a map
$\tilde D:\,\G_R\to\Aut(\C)$ is defined by $\tilde
D(\pi(\munt)):=D(\munt)$. This map $\tilde D$ now inherits the
representation property from $\tilde W$. \Halmos

\begin{theorem}[Spin-statistics connection]
$$F_\pm(\mc\otimes\phi)=\frac{1}{2}(1\pm F(\tilde {D}({{-1}})\mc\otimes\phi))$$
for all $\mc$ and all $\phi$.
\end{theorem}
{\it Proof.} 
For each $a\in S^2$ one has
$$\tilde W({{-1}})=J_aJ_{-a}=J_a\kappa
J_a\kappa^\dagger=J_a^2(\kappa^\dagger)^2 =k,$$
so
\begin{align*}
F(\mc\otimes\varphi)&=kF(\mc\otimes\varphi)k=\tilde W({{-1}})F(\mc\otimes\varphi)\tilde W({{-1}})\\&
=\tilde
W({{-1}})F(\mc\otimes\varphi)\tilde W({{-1}})^\dagger =\tilde
F({\tilde D}({{-1}})\mc\otimes\varphi).
\end{align*}\Halmos

\bigskip If, in particular, $\tilde{D}$ is irreducible with spin $s$,
then $\tilde{D}(-1)=e^{2\pi i s}$, so $F_-=0$ for integer $s$ and
$F_+=0$ for half-integer $s$.

\section{PCT-symmetry}

In order to justify the term ``modular P$_1$CT-symmetry'', one should 
show that this condition yields, at least in 1+3 dimensions, a full
PCT-operator in a base-independent fashion.

\begin{theorem}[PCT-symmetry] 
  There exists an antiunitary involution $\Theta$ with the properties
  
  (i) $J_aJ_bJ_c=\Theta$ for each right-handed orthogonal basis
  $(a,b,c)$ of $e_0^\perp$.

  (ii) There exists an antilinear involution $C$ such that 
  $$\Theta
  F(\mc\otimes\phi)\Theta=F(C\mc\otimes\overline\phi(-\,\cdot)).$$
\end{theorem}
{\it Proof.} Let $({a}',{b}',{c}')$ be a second right-handed
orthonormal base, and define $\Theta':=J_{a'}J_{b'}J_{c'}$.  Then it
follows from modular symmetry that
\begin{align*}
  \Theta'\Theta F(\mc\otimes\phi)\Omega
  &=\Theta'\Theta F(\mc\otimes\phi)\Theta\Theta'\Omega\\
  &=F(C_{a'}C_{b'}C_{c'}C_aC_bC_c\mc\otimes\phi)\Omega\\
  &=F(\tilde D(1)\mc\otimes\phi)\Omega\\
  &=F(\mc\otimes\phi)\Omega.
\end{align*}
Since $\Omega$ is cyclic, this implies the statement.\Halmos

\bigskip
If $(a,b,c)$ is right-handed and $(a',b',c')$ is left-handed, then
$\tilde D(1)$ has to be replaced by $\tilde D(-1)$ in the above
computation. Since $J_{-a}J_{-b}J_{-c}=\kappa J_aJ_bJ_c\kappa^\dagger$,
this is no surprise.

\section*{Conclusion}

Both the classical geometry and the fundamental quantum field
theoretic representations of the rotation group $SO(3)$ and its
universal covering group are based on reflection symmetries.  At the
classical level, the universal covering group $\G_R$ can be
constructed from P$_1$T-reflections. For a quantum field $F$ with
$\widetilde{SO(3)}$-symmetry, a class of antiunitary P$_1$CT-operators
exists that are fixed by the intrinsic structure of the respective
field. Along precisely the same lines of argument used for the
construction of ${\G}_R$, a covariant unitary representation $\tilde
W$ of ${\G}_{R}$ is constructed.  $\tilde W$ exhibits Pauli's
spin-statistics connection.

\subsection*{Acknowledgements}
This work has been supported by the Stichting Fundamenteel Onderzoek
der Materie and the Emmy-Noether programme of the Deutsche
Forschungsgemeinschaft. I would like to thank Professor Arlt, Klaus
Fredenhagen, and Reinhard Lorenzen for their critical comments and
help concerning this manuscript.

\begin{appendix}

\section*{Appendix. $SU(2)$ versus $\G_R$}  
  The isomorphism between the models $SU(2)$ and $\G_R$ of
  $\widetilde{SO(3)}$ can be described as follows.
  
  First recall the standard representation of $SU(2)$ on $\reals^3$.
  Denote by $\sigma_1,\dots,\sigma_3$ the Pauli matrices, and define
  $\hat x:=\sum_\nu x_\nu\sigma_\nu$, $x\in\reals^3$. For each $\nu$,
  the map $\hat x\mapsto\Ad(\pm i\sigma_\nu)\hat x$ is well known to
  implement the rotation $[e_\nu,\pi]$. Since the parity
  transformation $P$ is implemented by the map $\hat x\mapsto-\hat x$,
  one finds that for each $\nu$, the map $\hat x\mapsto
  -\Ad(\pm\sigma_\nu)\hat x$ implements the reflection $j_\nu$.  The
  determinants of the Pauli matrices equal $-1$, and all of them are
  involutions.
  
  Now one can define an isomorphism ${\mathfrak J}$ from $S^2$ onto
  the unitary matrices with determinant $-1$ by ${\mathfrak
    J}(a):=a\vec\sigma$. The products of pairs of unitary matrices
  with determinant $-1$ yield all of $SU(2)$.
\end{appendix}

\end{document}